\documentclass[12pt]{article}
\usepackage{amsmath,amssymb}
\usepackage{graphicx,epsfig}
 \def\cD{{\cal D}}
 \def\cS{{\cal S}}
\def\cF{{\cal F}}

\def\cL{{\cal L}} \def\cM{{\cal M}}
\def\cN{{\cal N}} 
 \def\cQ{{\cal Q}}

\def\beq{\begin{equation}}
\def\eeq{\end{equation}}
\def\bea{\begin{eqnarray}}
\def\eea{\end{eqnarray}}
\def\bet{\begin{tabular}}
\def\eet{\end{tabular}}
\def\bes{\begin{subequations}\bea}
\def\ees{\eea\end{subequations}}
\def\ol{\overline}

\def\a{\alpha}
\def\b{\beta}

\def\g{\gamma}

\def\e{\epsilon}
\def\ve{\varepsilon}
\def\l{\lambda}
\def\z{\zeta}
\def\bz{\ol{\z}}
\def\bl{\ol{\lambda}}

\begin{document}

\baselineskip 0.6cm

\begin{titlepage}

\vspace{5cm}

\begin{center}

{\LARGE \bf On orbifold theory and 
$N = 2$, $D=5$ \\ gauged supergravity}

\vspace{1cm}

Yin Lin

\vspace{1cm}



{\it Scuola Normale Superiore, Piazza dei Cavalieri 7,
I-56126 Pisa, Italy and
\\ Istituto Nazionale di Fisica Nucleare, Sezione di Pisa}

\end{center}

\vspace{1cm}

\begin{abstract}

We have studied the most general $N=2$ supergravity in five
dimensions in context with the orbifold theory based on $M_4
\times S^1/Z_2$. Various ways to treat the supersymmetry with
singular sources placed in orbifold fixed points were proposed 
in past. Supersymmetric branes were consistently introduced in a bulk where
a gauged supergravity was present. In
this paper we find that in the $N=2,D=5$ supergravity with general gauging,
 the possibility to obtain a supersymmetric brane
world is constrained. Imposing the compatibility of supersymmetry
transformation rules with the orbifold condition, we find the
necessary and sufficient condition to obtain supersymmetric branes and bulk
independently. We comment that the same condition guarantees naturally 
the presence of singular BPS solutions.

\end{abstract}

\vskip 20mm

\end{titlepage}

\newpage

\baselineskip 6 mm

\section{Introduction}

It's a very old idea that our four-dimensional world could be
embedded in a higher dimensional space-time. Recently more
attention concerned theories of large extra-dimensions, or brane
worlds, where the matter in our universe is confined to live on a $3+1$
dimensional hypersurface \cite{ADD} \cite{RS}. Phenomenological 
interest of brane-worlds is twofold. First of all, they introduce
extra-dimensions with a compactification scale testable by future experiments.
Secondly, a more fundamental motivation of these models was to provide a
solution to the hierarchy problem. In particular, in the
Randall-Sundrum steps \cite{RS}, a warped compactification has been explored.
In such models, the bulk space-time is an
$AdS_5$ or a slice of $AdS_5$ which induces on the branes an exponential decreasing warped factor.

Brane-world theories provide also alternative proposals
for understanding important problems in cosmology. In particular,
the dynamics of the bulk fields in a five dimensional model could
be the origin of a cosmological constant or slowly rolling ``quintessence'' scalar fields which are principal current candidates
of the dark energy. The smallness of the cosmological constant in
warped brane-world cosmology is presented as result of a self-tuning.
This idea advocates that, because of the warped factor, the
would-be high vacuum energy on our brane is converted simply in
the large curvature of extra-dimensions. This is the reason of the
apparent vanishing 4-D cosmological constant. It's however known
that the self-tuning domain wall suffers some kinds of instability
and the presence of naked singularity in the bulk. 
Problems arising due to the singularities have been point out in \cite{selftuning1} \cite{selftuning2}.
In some particular situations, singularity could be shielded by an additional 
source term, however this reintroduces a fine-tuning in the effective four dimensional theory \cite{selftuning2}.
Supersymmetry (SUSY) in the bulk provides a well known ingredient for
preventing above problems. From the five dimensional supergravity point of 
view, self-tuning is nothing but requiring a BPS brane world solution.
This is one of the motivations for us to reconsider the BPS construction
in a most general 5-D gauged supergravity with orbifold.

Brane-world arises naturally from gauged supergravities in five dimensions.
In the literature, gauged supergravity theories in five
dimensions were considered toward various directions, including or not localized interactions between bulk fields
and four-dimensional gauge theories on branes \cite{SUSYRS1} -- \cite{SUSYRS9}. To obtain
warped brane-world solution, the bulk supergravity must be modified by an
orbifold construction, for instance $S_1/Z_2$, likes the original step of Randall-Sundrum scenarios. 
Other works consider also the supersymmetry breaking in warped space due to boundary condition
\cite{SUSYB1} \cite{SUSYB2} \cite{SUSYB3} \cite{SUSYB4}. Recently, a general consistency 
study of higher dimensional theories with symmetry breaking defects at the boundaries is given \cite{SUSYB4}.
In this note we give a general framework of N=2 D=5 gauged supergravity in orbifold theory,
completing some of the previous constructions \cite{SUSYRS3} \cite{SUSYRS4} \cite{SUSYRS6}.
We find that some interesting constraints on the bulk supergravity would occur to give back it consistent
with orbifold condition.

From the five-dimensional point of view, brane-worlds are domain walls in space-time.
Introducing SUSY in space-times with ``fixed points'' looks at first sight as a severe problem.
In fact, in the orbifold theory, the Lorentz invariance in the direction orthogonal to the branes is broken.
Then we expect that SUSY, being the square root of translations, is broken as well.
The mechanism to restore the SUSY is to include on the branes the ``pull-back'' of suitable bulk fields in 
such a way that the supersymmetry variation of the branes compensates that of the bulk. 
This is the standard technique to incorporate the domain wall solution in supergravity theories.
We shall not concentrate on the detail of these possibilities \cite{SUSYRS1} -- \cite{SUSYRS9}. Our aim is to 
perform a general treatment on the consistency of the SUSY with the orbifold symmetry. 

The note is organized as follows: in the next section, we present an introduction of smooth N=2 D=5 gauged supergravity, describing 
all possible supermultiplets. In section 3 we introduce the orbifold condition and his connection with the scalar manifold.
In section 4 we will study in more detail the consistency between SUSY transformations and orbifold condition. We find
some constraints on the scalar potential of the supergravity theory. 
In section 5 a general treatment of sigma model coupled to 5-D gravity is considered.
The aim is to study the consistency of the brane action and the Israel junction conditions. In the end, in section 6, we
present an explicit supersymmetrization of the general gauged supergravity with orbifold
in absence of tensor supermultiplets .

\section{N=2, D=5 gauged supergravity}

The full couplings and gaugings of $D=5$ $N = 2$
supergravity were described in \cite{gaugetens} \cite{d5supergravity} and in the references therein. See also Appendix B for
some details of the full action helpful for our analysis.  
In this section, we shortly review the supermultiplets and scalar manifold present in this theory.

The most important fields of the supergravity theories are the components of the gravitational supermultiplet,
the only supermultiplet present in pure supergravity \cite{pure}:
$\left\{ e_{\mu}^a, \psi^{\a i}_\mu, A_\mu \right\}$. They are the graviton $e_{\mu}^a$, two gravitinos $\psi^{\a i}_\mu$
and a vector field $A_\mu$ (the graviphoton). To the pure supergravity, an arbitrary number of vector supermultiplets
$\left\{A_\mu,\l^i,\phi\right\}$ can be coupled.
The N=2 vector multiplet contains a vector field, an $SU_R(2)$
doublet of spin--$1/2$ fermions $\l^i$ and one real scalar field $\phi$.
One can introduce also matter supermultiplets:
$\left\{\z^A,q^X\right\}.$ 
A hypermultiplet contains a doublet of spin--$1/2$ fermions $\z^A$ and
four real scalars $q^X$ ($A=1,2$  $X=1,\ldots,4$). In detail, the supergravity theory 
with general matter coupling before the gauging is described by the following field content
\begin{equation}
\label{fields1}
\left\{ {e_\mu}^a, \psi^i_\mu, A_\mu^I, \l^{i x}, \z^A ,
\phi^{x}, q^u\right\},
\end{equation}
where an arbitrary number of vector supermultiplets ($n_V$) and hypermultiplets ($n_H$) is introduced. 
The index A now runs from 1 to $2n_H$. The scalars $\phi^{x}, q^u$
are coordinates of a scalar manifold $\cM$ which is a 
direct product of a so called very special manifold and a
quaternionic manifold:
\begin{equation}
\cM = \cM_V \otimes \cM_H,
\end{equation}
with  $\phi^{x} \in \cM_V$, $q^u \in \cM_H$. $u=1,\ldots, 4n_H$ and $x=1,\ldots, n_V$ are the curved indexes labeling the
coordinates of $\cM$. 
In (\ref{fields1}), $\l^{x i}$ has two kinds of index, the components labeled by $x$ transform as a vector under the tangent
space group $SO(n_V)$ of vector scalar manifold $\cM_V$ while $i$ is an $SU_R(2)$ rotation index. 
The index of $\z^A$ is flat.

The scalar manifold $\cM$ admits in general a large group of isometries $G$ and some of its subgroups $K$,
as well as the $SU_R(2)$ group, can be gauged using the vector fields present in the theory. One obtains in this way the so called 
gauged supergravity. An important property of the gauged supergravities is the presence of a potential for 
scalar fields. This is a necessary condition to the realization of AdS vacua and BPS solutions \cite{hyperBPS}. 

The non abelian gauging in a theory where only abelian vectors are present is problematic \cite{gaugetens}. The reason is that if 
the gauge group $K$ is non abelian, some spectator vector fields could be charged under $K$. Since before gauging all vectors 
are abelian, one has an inconsistency between the abelian gauge symmetry and the non abelian gauging. To solve this problem,
the SUSY imposes that the K-charged spectator vectors have to be dualized to {\it self-dual} two-form fields.
As a result, the 5D gauged supergravity may contain also tensor supermultiplets:
$\left\{B_{\mu\nu},\l^i,\phi \right\}$, here again an $SU_R(2)$ doublet of spin--$1/2$ fermions and one
real scalar field has been required.
The tensor field satisfies the self-duality condition:
\beq
\label{tensor1}
B_{\mu \nu} \propto \varepsilon _{\mu \nu \rho \sigma\tau } \partial ^{[ \rho} B ^{\sigma \tau ]}
\eeq

to guarantees the same degree of freedom of a vector $A_{\mu}$.
 The existence of tensor multiplets
does not change the scalar content of the theory but the form of the scalar potential. 

Now we review some important characteristics of the general $SU_R(2)$ gauging.
In the absence of hypermultiplets, the total global symmetry group factorizes
into $SU(2)_{R}\times G$, as a consequence of the  $SU(2)_{R}$-invariance
of the scalar fields belonging to the vector multiplets. However, in general 
matter coupled supergravity theories, the R-symmetry group is 
non-trivially embedded into the quaternionic global symmetry group. In fact, in a quaternionic manifold, 
the holonomy group is $USp(2n_H)$ times $USp(2)\simeq SU(2)$. The R-symmetry group is identified with the $USp(2)\simeq SU(2)$
contained in the holonomy group of the quaternionic manifold. Since $USp(2)$ is evaluated on scalar manifold, 
when the whole  $SU(2)_{R}$ symmetry is gauged, a rotation of $SU(2)_{R}$ may depend itself on scalar fields.

Five dimensions supergravity has a Chern--Simons term, a peculiar feature 
of odd-dimensional space--times. This term characterizes the ``very special'' 
manifold described by the scalars of vector multiplets. The gauging of the theory is then
quite different from 4D gauged supergravity where the scalar manifold is special K$\ddot{a}$hler.

\section{Scalar manifold and $\mathbf Z_2$}

Now we can incorporate the supergravity theory just discussed in an orbifold construction: $M_5=M_4  \otimes S_1/Z_2$.
Identifying the smooth circle $S_1$ with the fifth direction $r$,
all fields of the theory are considered periodic in r.
More precisely, if the radius of the circle is $R$, all fields obey $\Phi (r)=\Phi
  (r +2\pi R)$. In fact alternative theories were studied in this regard to get
spontaneous supersymmetry breaking. For instance, \cite{SUSYB1} 
\cite{SUSYB3} one can consider also the possibility to have a flipping in the boundary condition for fermionic fields, i.e. 
$\Psi (r)=-\Psi(r +2\pi R)$. We do not consider this possibility.
The next step to get the orbifold theory on 5D gauged supergravity
is to impose the $\mathbf Z_2$ symmetry which acts by $r \to -r$ on the fields.
On the bosonic fields, the action of $\mathbf{Z_2}$ is
the standard parity transformation on the fifth dimension:
$$\Phi(r)=\mathbf{P}(\Phi)\Phi(-r)=\pm \Phi(-r).$$
The fermions  $\psi^{\a i}_\mu, \l^i$, as well as
the supersymmetry parameters $\ve^i$, are symplectic (see Appendix A) carrying an $SU(2)_R$ index, their
parity transformation needs to include an $SU(2)_R$ rotation:
\begin{eqnarray}
\lambda ^i(r)= \mathbf{P}(\lambda )\gamma _5 M(q)^i{}_j \lambda ^j(-r)\,,\\
\psi^i_{\mu}(r)= \mathbf{P}(\psi )\gamma _5 M(q)^i{}_j \psi ^j(-r)\,,\\
\ve ^i(r)=  \mathbf{P}(\ve )\gamma _5 M^i(q){}_j \ve ^j(-r)\,,
\end{eqnarray}
where
\begin{equation}
  M^i{}_j=m_1(q) (\sigma _1)^i{}_j + m_2(q) (\sigma _2)^i{}_j + m_3(q) (\sigma
  _3)^i{}_j\,,
\label{Mherm}
\end{equation}
$\mbox{with}\ m_1,\, m_2,\,m_3\in$ real functions of $q$\footnote{In the case of minimal gauged supergravities with an abelian gauging in a specific direction, the orbifold rotation $M^i{}_j$ must be constant.} and $\mathbf{P}(\Psi)=\pm$, $\Psi\equiv(\lambda,\psi,\ve)$. 
Due to the parity property of $M$: $M^2=1$, 
the coefficients $m_i$ must form a vector with unit norm.
The news here with respect to the previous works is that we introduce a dependence of the rotation coefficients on the scalar fields. 
The reason is that, in the general gauged 5D supergravity under consideration,
the whole $SU(2)_R$ group has been gauged rather than its abelian subgroup $U_R(1)$.
On the other side, as already discussed in the previous section, $SU(2)_R$ is just identified with the subgroup $USp(2)$
of the holonomy group
of the quaternionic manifold. As will also be shown explicitly in the following, $m_i$ depends in fact only on scalars $q$.

Only the even field will survive on fixed points of the orbifold where
we will put the branes. What we are going to do in the next section is to assign the correct 
parity properties to the fields in order to keep the bulk action even and all
supersymmetric transformations consistent. This is a fundamental step also for a successful supersymmetrization
of the branes themselves. As a first and obvious requirement, we shall assign a negative parity to the $r$ coordinate:
$\mathbf{P} (\partial_5)=-1$. Furthermore, $\mathbf{P} (e_m {}^{\tilde{a}})=1$ because the brane 
action for the 4-dimensional hypersurface is multiplied by the determinant of the vierbein $e_m {}^{\tilde{a}}$.

\section{Susy transformations and $\mathbf Z_2$}

For clearness, we recall the full SUSY transformation rules for
the general gauged $N=2$ $D=5$ supergravity (for the notation, see Appendix B):

\bea
\label{susytransform1}
\delta_{\ve} e^a_{\mu} &=&
\frac{1}{2} \bar{\ve}^i \g^a \psi_{\mu i},\\
\delta_{\ve} \psi_{\mu i} &=&
D(\widehat{\omega}) \ve_i + \frac{i}{4\sqrt{6}} h_{I}  e^a_{\mu}
\left( \g_{abc}\ve_i - 4 \eta_{ab} \g_c \ve_i\right)
\widehat{H}^{bc\,I} - \delta_\ve q^X {\omega_{X\,i}}^j \psi_{\mu j}
+   \nonumber \\
&-& \frac{1}{12} e^a_{\mu} \g_{ab} \ve^j \;
\bl^{x}_i \g^b \l_j^{x} + \frac{1}{48} e^a_{\mu} \g_{abc}
\ve^j \;  \bl_i^{x} \g^{bc} \l_j^{x} +
\frac{1}{6} e^a_{\mu} \ve^j \;  \bl_i^{x} \g_a \l_j^{x}
+ \nonumber \\
&-& \frac{1}{12} e^a_{\mu} \g^b\ve^j \;
\bl_i^{x} \g_{ab} \l_j^{x}  + \frac{1}{8} e^a_{\mu}
\g^{bc} \ve_i \; \bz_A \g_{abc} \z^A
+\frac{i}{\sqrt{6}} g_R \, e^a_{\mu} \g_a \ve^j P_{ij}, \\
\delta_{\ve}
\phi^{x} &=&  \frac{i}{2} \bar{\ve}^i \l_i^{x} \\
\delta_{\ve} A^I_{\mu} &=&  - \frac{1}{2} e^a_{\mu} \,  \bar{\ve}^i \g_a
\l_i^{x}\, h_{x}^{I} + \frac{i\sqrt{6}}{4}
\ol{\psi}^i_{\mu} \ve_i \, h^{I},\\
\delta_{\ve} \l_i^{x} &=& -
\frac{i}{2} \g^a \ve_i \, \widehat{D}_a \phi^{x}
- \delta_\ve \phi^{y} {\Omega_{y}}^{xz} \l^{z}_i -
\delta_\ve q^X \omega_{X\,i}{}^j \l^{x}_j +\frac{1}{4}
h_{I}^{x} \g^{ab} \ve_i \,  \widehat{H}_{ab}^{I}  +
\nonumber \\
&-& \frac{i}{4\sqrt{6}} {T^{x}}_{yz} \left[ -3
\ve^j \, \bl^{y}_i \l_j^{z} + \g_a\ve^j \,
\bl^{y}_i \g^a \l_j^{z} + \frac{1}{2} \g_{ab}
\ve^j  \, \bl^{y}_i \g^{ab} \l_j^{z} \right]+ \\
&+& g_R \ve^j P^{x}_{ij} + g  W^{x} \ve_i,  \nonumber \\
\delta_{\ve} B^M_{\mu \nu} &=& 2D_{[ \mu}\delta_{\ve} A^M_{\nu ]} + \frac{i}{8} g  e^b_{\mu} e^a_{\nu} \,
\bar{\ve}^i \g_{ab} \l_i^{x} \, h_{N} \Omega^{MN} +
g \frac{\sqrt{6}}{8} e^a_{\mu} \, \ol{\psi}^i_{\mu} \g_a \ve_i \,
\Omega^{MN} h_N, \label{susytensors}\\
\delta_{\ve} q^X &=& - i \bar{\ve}^i \z^A \, f_{iA}^X, \\
\delta_\ve \z^A &=& - \frac{i}{2}\g^a \ve^i \widehat{D}_a q^X \,
f_{iX}^A - \delta_\ve q^X {\omega_{X\,B}}^A \z^B +  g  \ve^i \cN_{i}^A.
\eea

We begin to choose the components of the gravitino in the directions transverse to `5', $\psi _m $, to be even.
We will return to this point when we study the supersymmetry on the branes.
First let us concentrate on the ungauged part of the SUSY transformation.
Only the following parity assignments are then consistent:
\bea
\mathbf{P} (e_m {}^{\tilde{a}})=\mathbf{P} ( e_5{}^5)=1\,,\qquad
\mathbf{P}  ( A_5^I)=1\,,\nonumber \\
\mathbf{P}(q)= \mathbf{P}(\phi)=1 \,, \qquad \mathbf{P} (\psi_m)=\mathbf{P} (\epsilon )=1\,,\\
\mathbf{P} (e_5{}^{\tilde{a}})=\mathbf{P}
(e_m {}^5)=-1\,, \qquad \mathbf{P} (A^I_m)= \mathbf{P}(\Lambda^I)= -1\,, \nonumber\\
\mathbf{P} (\psi _5)=\mathbf{P}
(\lambda)=\mathbf{P}(\z)=-1.  \label{EvenOdd}
\eea

What is left over are the parity properties of the tensor fields which must be considered apart.
As we have already seen, the tensor fields in $N=2$ $D=5$ gauged supergravity are self-dual:
\beq
\label{tensor}
B^M _{\mu\nu}= \frac {1}{m_T} \varepsilon _{\mu \nu \rho \sigma\tau} \partial ^{[\rho} B ^{M\sigma\tau]},
\eeq
where $M = 1, \ldots, n_T$ labels the tensor multiplets and $m_T$ has dimension of a mass. 
A naive statement from the SUSY transformation (\ref{susytensors}) provides the choice $$\mathbf{P} ( B_{m 5}) = -\mathbf{P} 
(B_{m n})=1.$$
However it is easy to point out that this choice is not consistent with the condition (\ref{tensor}).
To show it, we can consider the component $mn$ of (\ref{tensor})
\beq
\label{self}
B_{mn}= \frac{1}{m_T} \left( \varepsilon _{mn5pq}\partial^{[5}
 B^{pq]}+ \varepsilon _{mnl5q}\partial^{[l} B^{5q]} +\varepsilon _{mnlp5}\partial^{[l} B^{p5]} \right)
\eeq
Self-duality condition (\ref{self}) mixes the $B_{m 5}$ components with the $B_{m n}$ ones.
Since $\mathbf{P}(\partial^l)=1$ and $\mathbf{P}(\partial^5)=-1$, (\ref{self}) alone cannot be realized with 
defined parity assignment of $B_{\mu \nu}$. The only way to get around this problem is to introduce 
parity property also for the mass of tensors $m_T$: $$m_T \longrightarrow m_T \varepsilon (r)$$ where $\varepsilon (r)$
is the step function. We then conclude that {\it tensor fields are consistent with the orbifold theory only if they have an odd mass.}

It remains to analyze the consistency of the SUSY transformation rules containing the gauge constant $g_R$.
The two relevant terms in this case appear in the trasformation of the gravitino and of $\l_i^{x}$:
\beq
\frac{i}{\sqrt{6}} g_R \, e^a_{\mu} \g_a \ve^j P_{ij}\,, \qquad g_R \ve^j P^{x}_{ij}.
\eeq 
Taking the parity of both sides, we obtain the following two conditions:
\bea
\label{cond1}
 M^j{}_iP^k{}_{j}\epsilon _k=-\mathbf{P} (g_R) P^j{}_i M^k{}_j\epsilon _k\,,\\
\label{cond2}
 M^j{}_iP^{x k}{}_{j}\epsilon _k=-\mathbf{P} (g_R) P^{x j}{}_i M^k{}_j\epsilon _k\,.
\label{MPPM}
\eea
where we allowed a possible flipping of the sign in $g_R$ when it goes through
the origin of the fifth dimension. From the property of very special geometry 
((\ref{relation}) in Appendix B), we have the relation $P^{x}_{i j}= -\sqrt{\frac32} 
P^{,x}_{i j}$ where $^{,x}$ denotes $\partial_{\phi^x}$.
Then confronting the derivative of (\ref{cond1}) with (\ref{cond2}),
the $SU(2)_R$ projector $M$ can be writing in the following form:
$$M_{ij}=f(\Phi)M^0 _{ij}$$
with $M^0 _{ij}$ independent on $\phi^{x}$ and $f$ a scalar function with $f^{2}=1$. 
Note that it is not restrictive to impose $f \equiv 1$. In this way M does not
depend anymore on $\phi^{x}$.

The prepotential $P_{ij}$, as well as $M_{ij}$, is often written in the base of the Pauli matrices $\sigma^s$
in the following: $$P_{ij}=i(P^s \sigma^s)_{ij}\,, \qquad M_{ij}=i(M^s \sigma^s)_{ij}. $$
The SUSY transformation laws are $\mathbf{Z_2}$ invariant if we may decompose
\beq
\label{decomp}
g_R \bar{P}= g_R^{(1)} \varepsilon (r) \bar{P}^{\parallel}+ g_R^{(2)} \bar{P}^{\perp},
\eeq
where $\bar{P}^{\parallel}$ is parallel to $\bar{M}$ while $\bar{P}^{\perp}$ orthogonal to $\bar{M}$.
Or equivalently, $P^{\parallel}_{ij}$ commutes with $M_{ij}$ and $P^{\perp}_{ij}$ anticommutes with $M_{ij}$.
In the former case, the gauge coupling is proportional to the step function $\varepsilon(r)$.

The different features for the parallel and the orthogonal components are well known \cite{SUSYRS4} \cite{SUSYB1}.
The presence of $\bar{P}^{\perp}$ breaks the BPS domain wall solutions.
The fact is one would not solve Killing equations everywhere in the bulk because
odd fields are discontinuous in fixed points of the orbifold. 
To restore the SUSY, the techniques used are to add 
new terms proportional to $\delta$, delta of Dirac, in SUSY transformation laws and consequently
a brane action to guarantee the closure of $N=2$ supersymmetry. 
However in \cite{SUSYRS4} it has been pointed out that this mechanism does work only in the absence of hypermultiplets. 
The main reason is that in presence of a general scalar manifold, the equations of motion 
of the scalar fields destroy the desired closure of the SUSY algebra.
From the previous discussion, since this note considers general matter coupled supergravity,
we renounce to the possibility to have an orthogonal component in (\ref{decomp}) therein.

Now let us return to the consistency conditions (\ref{cond1}) and (\ref{cond2}). Since we
consider only the parallel component of the prepotential, $P_{ij}$ must be
equal to $M_{ij}$ times a scalar function:
\beq
\label{PandM}
P_{ij}=A(q,\phi)M_{ij}(q).
\eeq
In other words, we can decompose the prepotential $P_{ij}$ in its norm $W(q, \phi)$, which is precisely 
the superpotential, multiplied by its $SU_R(2)$ phase $Q^s$:
\beq
\label{result}
P^s=W(q, \phi)Q^s, \qquad \textrm{with the constraint} \qquad \partial_{\phi}Q^s=0.
\eeq
The most important consequence of this result (\ref{result}) is, in agreement with \cite{hyperBPS}, that
the scalar potential $V(q, \phi)$ can then be written in a form that has been put forward for gravitational stability
neglecting the tensor contribution:
\beq 
\label{potstable} 
{V}= - 6 W^2+\dfrac92 g^{\Lambda\Sigma}\partial_\Lambda
 W\partial_\Sigma W\,.
\eeq
$g_{\Lambda \Sigma}$ represents the metric of the complete scalar manifold.
$N=2$, $D=5$ gauged supergravities with scalar potential given in the stable form had already been analyzed 
to establish supersymmetric domain walls flow in these theories \cite{flux} \cite{hyperBPS}. 
In particular, in \cite{hyperBPS}, the authors
impose the condition (\ref{result}) by hand in order to obtain stable form of the scalar potential and 
consequently analyze supersymmetric domain wall solutions.
Our result, on the contrary, points out that: 
\begin{quote}
\emph{the consistency between the orbifold conditions and the SUSY transformations yields automatically the stable form of the scalar potential (\ref{potstable}).} 
\end{quote} 
Furthermore, in the following sections we will show that the stable form of the scalar potential is also necessary 
to successful supersymmetrization of the full brane-bulk system. 

\section{Brane action}

The gauge coupling multiplied by a step function $g_R\varepsilon(r)$ introduces a singularity in the SUSY transformations law.
 The bulk supergravity will not longer be supersymmetric and its variation
must be proportional to the delta function centered on the fixed points of the orbifold.
It's then necessary to introduce a brane action to cancel this variation.
In this section, we will show how the stable form imposed on the superpotential is required to
obtain a consistent brane action\footnote{Tensor multiplets give otherwise contribution to the scalar potential which destroys
the stable form and we will not consider them here.}. The brane action is also accountable to the matching
of various slides of the bulk supergravity in the orbifold background.
To understand all these points, let us concentrate on the scalar part of the bulk action:
\bea
\label{scalaraction}
e^{-1}\cL_{scalar} &=& \frac{1}{2} R -
\dfrac{1}{2} g_{XY} \partial_\mu q^X \partial^\mu q^Y - \dfrac{1}{2} g_{xy} \partial_\mu \phi^{x} 
\partial^\mu \phi^{y} - V(\phi,q) \nonumber\\
&=& \frac{1}{2} R -
\dfrac{1}{2} g_{\Lambda \Sigma} \partial_{\mu} \Phi^{\Lambda} \partial^{\mu} \Phi^{\Sigma} - V(\phi,q),
\eea
where $V(\Phi)$ is the scalar potential expressed in his stable form (\ref{potstable}).
Remember that the metric $g_{\Lambda \Sigma}$ represents the complete scalar manifold and we assume the 
tensor contributions to the scalar potential to vanish.
Notice that tensor multiplets in general give contributions to the scalar
potential that destroy its stable form.

The theory described by (\ref{scalaraction}) can be seen as a Sigma model coupled to gravity 
which is a generalization of scalar-tensor gravity.
Domain wall solutions in scalar-gravity are quite interesting issues \cite{selftuning3} \cite{flux}.
Here we will examine the Einstein theory in the Sigma model case, because the metric $g^{\Lambda \Sigma}$ in (\ref{scalaraction})
depends itself on the scalar fields. 

We look for static solutions with a metric ansatz which depends only on the transverse direction of the brane
and is Minkowskian flat on the brane. 
\beq
	\label{metric}
ds^2=e^{-A(r)} dx^2_4 + dr^2.
\eeq
The equations of motion which follow from the action (\ref{scalaraction})
with the ansatz (\ref{metric}) are:
\bea
	\label{ein1}
\frac32A''(r)  = g_{\Lambda \Sigma} \Phi'^{\Lambda}(r) \Phi'^{\Sigma}(r),
\\
	\label{ein2}
\frac32 A'(r)^2 =\frac12 g_{\Lambda \Sigma} \Phi'^{\Lambda}(r) \Phi'^{\Sigma}(r) - V[\Phi(r)],
\\
	\label{ein3}
(g_{\Lambda \Sigma} \Phi'^{\Sigma}(r))'-A'(r)g_{\Lambda \Sigma} \Phi'^{\Sigma}(r) =
2\frac{\partial V[\Phi]}{\partial\Phi}
\eea
where the primes denote derivatives with respect to $r$.
Since the potential $ V[\Phi]$ is given in terms of the superpotential $W$ in the stable form,
it is very easy to see the general solution of the system (\ref{ein1})-(\ref{ein3}):
\begin{eqnarray}
	\label{phi'}
\Phi'^{\Lambda}(r)&=& 3g^{\Lambda \Sigma} \frac{\partial W[\Phi(r)]}{\partial\Phi^{\Sigma}(r)},
\\
	\label{A'}
A'(y)&=&2W[\Phi(r)] .
\end{eqnarray}
We can introduce now a general brane action at a fixed point $r=r_0$
\beq
\label{tension}
\int d^{5}x\,\sqrt{-g}\,\,f[\Phi]\,\delta (r-r_0).
\eeq
The equations of the motion will be modified and in order to construct continuous solutions in the bulk with now two slides
separated by the brane, they need to respect a set of supplementary
boundary conditions at the brane:
\bea
\label{junction}
&&
\Delta \Phi'^{\Lambda}(r)=
3\Delta g^{\Lambda \Sigma} \frac{\partial W[\Phi(r)]}{\partial\Phi^{\Sigma}(r)} =
g^{\Lambda \Sigma} \frac{\partial f[\Phi(y)]}{\partial\Phi^{\Sigma}(r)}\Big|_{r=r_0},
\nonumber \\
&&
\Delta A' =
3\Delta W[\Phi(r)]={f[\Phi(r)]}\Big|_{r=r_0},
\label{bc} \eea
where $\Delta F$ indicates the jump of a discontinuous function $F$ at
$r=r_0$, $F(r_0^+)-F(r_0^-)$. These conditions are known as Israel junction conditions.
Obviously $\phi(r)$ and $A(r)$ are automatically continuous across the boundary
in the orbifold background. From (\ref{junction}), it is easy to check that in order to
be consistent with the boundary condition for any $r_0$, the coefficient
$f[\Phi]$ in front of the determinant of the 4-dimensional metric in (\ref{tension}) must be exactly 
three times the superpotential $W$.
Since our background is $\mathbf{S^1}/ \mathbf{Z_2}$, there are 2 fixed points at $r=0$ and $r=\pi R$, the brane action is
\beq
\label{Sbrane1}
S_{brane} \supset -\int d^5x \left( \delta (r)-\delta
(r-\pi R)\right) e_{(4)}3W.
\eeq
where $e_{(4)}$ is the determinant of the vierbein $\e_m^{\tilde{a}}$.
Now the importance of the stable form of the potential in
the orbifold theory appears clear. In the absence of the condition obtained in (\ref{result}),
it's impossible to introduce a brane consistent with the boundary condition for any radius R of the extra-dimension. 
From the supergravity point of view, this conclusion is mainly due to the fact that junction conditions 
are also conditions for the existence of BPS brane-world solutions \cite{flux}.

\section{Supersymmetrize brane and bulk}

In this section, we extend the method explained in \cite{SUSYRS4} to supersymmetrize 5-dimensional supergravity
coupled with vector multiplets and hypermultiplets with general gauging. As a result,
we find that the assignment of the parity of various fields is fundamental in order to
supersymmetrize branes independently to bulk supergravity.
The full action is described in Appendix B.
The method consists in replacing the odd gauge coupling constant $g_R \varepsilon (r)$
by a new smooth scalar field $G(r)$ which is
invariant under SUSY transformations. One then introduces a four form $A _{\mu \nu \rho \sigma }$
as a Lagrangian multiplier to impose $G(r)$ to be everywhere constant on shell:
\beq
\label{SA}
  S_{A}=\frac{1}{4!} \int d^5 x \varepsilon ^{\mu \nu \rho \sigma\tau} A _{\mu \nu \rho \sigma} \partial _{\tau} G \,.
\eeq
Varying the action with respect to $A_{\mu \nu \rho \sigma }$ we obtain the usual supergravity action
in the bulk. 
The advantage of this formulation arises when we wish to supersymmetrize the branes and the bulk independently.
The replacement $g \to G$ breaks the original $5D$ SUSY, but the SUSY variation of
(\ref{SA}) may restore it. 
\bea 
\label{deltaS}
 e^{-1}  \delta(\epsilon) {\cL}_{5d}
&=&\left[ - \overline{\psi }{}_\mu^i\gamma ^{\mu \nu\rho  }\epsilon
  ^j A^{I}_\nu P_{Iij}-\frac{i \sqrt{6}}{8} \overline{\psi
}{}_\mu ^i\gamma ^{\mu \nu }\epsilon
  ^j P_{ij}+ \right. \nonumber \\
&+& \left. 2 \overline{\lambda }{}^i_x
  \gamma ^\nu \epsilon ^j P^x_{ij} -2\overline{\ve^i} \gamma ^{\nu}\z^{A} \cN_{iA} \right] \partial _\nu G\, \nonumber\\
&=& B^{\nu}\partial _\nu G.
\eea
This variation is proportional to $\partial _\nu G$ since when $g_R$ is constant, the gauged action is supersymmetric.
If we define the SUSY transformation of $A_{\mu \nu \rho \sigma }$ as 
\beq
 \delta(\epsilon) \frac{1}{4!} \varepsilon ^{\mu \nu \rho \sigma \tau } A_{\nu \rho \sigma \tau}=-B^{\mu}\,,
\eeq
the whole bulk remains supersymmetric and the SUSY algebra will be realized on-shell.

The boundary action consists of two parts. The first one is (\ref{Sbrane1}),
already considered in the previous section.
The second part is the induced action (\ref{SA}) on the fixed four-dimensional
sub-spaces of the orbifold. In summary, we have
\beq
\label{Sbrane}
S_{brane}=-2 G(r) \int d^5x\,\left( \delta (r)-\delta
(r- 2 \pi r_0)\right)
  \left( e_{(4)} 3W +\frac{1}{4!} \varepsilon ^{mnpq} A _{mnpq} \right) \,.
\eeq
Let's observe that the second part of the brane action (\ref{Sbrane})
is consistent with the parity assignment of $G$:
$$\mathbf{P}(G)=-1 \qquad \longrightarrow \qquad \mathbf{P}(A_{mnpq})=+1.$$
Having introduced the brane action, the field equation of $G(r)$ becomes:
\beq
  \partial _5 G(r)= 2 g_R \left( \delta (r)-\delta (r-2 \pi r_0 )\right),
\eeq
which implies $G(r)=g_R \varepsilon (r)$. To check the SUSY invariance of the brane action we use the formula 
$$\delta_{\ve}W=W_{,x}\delta_{\ve}\phi^x +W_{,u}\delta_{\ve}q^u $$ 
and the variation of $A_{mnpq}$ from (\ref{deltaS}).
\bea
  \delta_{\ve}  S_{brane}= &-& 3g\int d^5x \left( \delta (r)-\delta (r-2 \pi r_0 )\right)
 \nonumber\\ &\times&  \left[ W e_{(4)}   \overline \epsilon ^i \gamma^{\tilde{a}} e_{\tilde{a}}^m 
\left(\psi _{mi} - i \gamma _5 Q_{ij}\psi _m^j \right)+ W_{,x}
  \overline \epsilon  ^i i\left( \lambda _i^x +i \gamma _5 Q_{ij}
  \lambda ^{xj}
\right) \right.\label{deltaSbr1}\\
&+& \left. \overline \epsilon ^i \z^A \left(-2iW_{,u} f_{iuA}-\frac43 \gamma_5 \cN_{iA} \right) \right] \nonumber.
 \label{deltaSbr2}
\eea
where $Q_{ij}$ will be replaced by $M_{ij}$, see (\ref{PandM}) and (\ref{result}). The terms in the second line of (\ref{deltaSbr1}) 
come from the vector multiplets and are result has been obtained in \cite{SUSYRS4}. 
The only difference, in our case, is that $Q_{ij}$ now depends on scalars $q^u$ of the quaternionic
manifold. The terms in the third line of (\ref{deltaSbr2}) are the contribution of hypermultiplets. 
For the parity assignment (\ref{EvenOdd}), the following fields or combination of fields
\beq
\label{fieldsbrane}
\psi_{mi}^- = \psi _{mi} - i \gamma _5 M_{ij}\psi _m^j \,, \qquad
(\lambda_i^x)^- =  \lambda _i^x +i \gamma _5 M_{ij} \lambda ^{xj} \,, \qquad \z^A \,,
\eeq
have parity odd and then they vanishes on the branes. In principle, there is a second possibility of 
parity assignment changing simultaneously the parity of the SUSY transformation parameter $\epsilon$
and of $\psi_{\mu}^i$. In theories where only vector multiplets are present, the second choice is equivalent of the first one
because we can add an additional minus sign when we identify $Q_{ij}$ with $M_{ij}$. But when we consider also the hypermultiplets, a negative assignment for $\epsilon$ implies a positive one 
for $\z^A$ and the variation (\ref{deltaSbr1}) does not vanish anymore. We check that in order to obtaining
supersymmetric branes, (\ref{EvenOdd}) is the only possible assignment of parity. 

\section{Conclusion}

In summary, this note reviews some salient aspects concerning the combination of gauged supergravity theories 
with recent well promising orbifold theories. $5D$ supergravity realization of phenomenologically testable brane-world models
is a very attractive issue. Many mechanisms were introduced in order to implement supersymmetry in space-time with singular sources.
Although in the last few years, solutions in this regard were obtained in different way, it still remains an interesting stage
to explore some general connection between 5D gauged supergravity and orbifold theory. Motivated by this intent,
our goal has focused on the auto-consistency of the SUSY algebra, when matter fields are present in the bulk, with the orbifold 
construction and with the brane-world solution. An important requirement in searching singular BPS solutions in this theory
is the stable form of the scalar potential. In fact, in general matter coupled supergravity, a
gravitational stable potential is not automatic. In a different way, we find that orbifold theory can guarantee a stable form of scalar 
potential by the orbifold projection. As a result, the existence of SUSY brane-world solutions can emerge naturally.
Furthermore, these solutions are important candidates also for cosmological scalar-tensor gravity in context of a
more fundamental background. In this perspective, it's interesting to find a specific model where 
hypermultiplets lead to a supersymmetric brane-world solution with warped compactification and we
hope to come back to this point in near future.

\vskip 1truecm

\paragraph{Acknowledgments.}
We are glad to thank Mario Tonin for helpful conversations and for pointing out some misprints.
We are particularly grateful to Maria Laura Alciati and Luca Scarabello for their interest and encouragement in our work.
This work is partially supported by the EU under TMR contract HPRN-CT-2000-00148.

\vskip 1truecm
\section*{Appendix A: Notations}
The metric is $(-++++)$, and $[ab]$ denotes antisymmetrization $\frac{[ab-ba]}2$. We use the following indexes
\begin{eqnarray}
 \mu  & 0,\ldots ,3,5 & \mbox{local spacetime}     \nonumber\\
 m & 0,\ldots ,3 & \mbox{4D local spacetime} \nonumber\\
 a & 0,\ldots ,3,5 & \mbox{tangent spacetime}     \nonumber\\
 \tilde{a} & 0,\ldots ,3 & \mbox{tangent spacetime in 4 dimensions}\nonumber
\label{indices}
\end{eqnarray}
The Levi--Civita tensor satisfies
\begin{equation}
  \varepsilon _{abcde}\varepsilon ^{abcde}=-5!\,,\qquad
  \varepsilon ^{\mu \nu \rho \sigma\tau }=ee^\mu _ae^\nu _b\cdots e^\tau _e\varepsilon
^{abcde}\,.
 \label{LeviCivita}
\end{equation}
The five--dimensional ${(\g^a)_\a}^\b$ matrices satisfy the Dirac algebra
$$
\{ \g^a, \g^b \} = 2 \, \eta^{ab}$$.
$\g ^{a_1 \ldots a_n}$ means the antisymmetrized product of $n$ $\g$
matrices with weight one: $\g^{[a_1} \ldots \g^{a_n]}.$

Symplectic--Majorana spinors carrying the $USp(2)$
doublet index $i=1,2$ which is raised and lowered with
the invariant $USp(2)$ tensor $\e_{12}= \e^{12} = 1$, in NW-SE convention:
\begin{equation}
  X^i=\varepsilon ^{ij}X_j\,,\qquad X_i=X^j\varepsilon _{ji}\,.
\label{NWSEconv}
\end{equation}
The symplectic--Majorana condition on a generic spinor $\l_{\a i}$
reads
\beq
\ol{\l}^i \equiv \l_i^\dagger \g^0 = \left(\l^i\right)^T C,
\eeq
where $\ol{\l}$ is the usual Dirac conjugate and $C$ is the charge
conjugation matrix satisfying
$C^T = -C = C^{-1}$.

\section*{Appendix B: N=2 D=5 gauged supergravity}
\label{AppendixB}

\begin{eqnarray}\label{Lagrange2}
e^{-1}\mathcal{L}&=& -\frac{1}{2} R(\omega) - \frac{1}{2}
{\bar{\psi}}_{\mu}^{i}\gamma^{\mu\nu\rho}(\nabla_{\nu}(\omega)\psi_{\rho i}+\cD q^u \omega _{uij} 
\psi_{\rho}^j+g_R A_{\nu}^I P_{Iij}\psi_{\rho}^j) \nonumber \\
&& -\frac{1}{2}{\bar{\lambda}}^{i}_{x}\gamma^{\mu} \left[ \nabla_{\mu}(\omega) \l^{x}_{i} +g_R  A^I L_I{}^{xy}(\phi)
\l_{yi} \right. \nonumber \\
&& + \left. \Gamma^{x}_{yz}(\cD_{\mu} \phi^{y} ) \l^{z}_{i}+
\cD_{\mu} q^u \omega_{ui}{}^j \l^{x}_{j}+g_R A_{\mu}^I P_{Iij} \l^{xj}\right] + \nonumber \\
&& - \bar\z_A \gamma ^{\mu}\nabla_{\mu}(\omega)\z^A+ 
\bar\z_A \gamma ^{\mu} \left( \cD _{\mu}q^u \Delta_{uB}{ }^{A}+g_R A^{I}_{\mu} \omega_{IB}{ }^A \right) \z^B \nonumber\\
&& + \frac{e^{-1}}{6\sqrt{6}}C_{IJK}\varepsilon^{\mu\nu\rho\sigma\lambda}
\left\{ F_{\mu\nu}^{I}F_{\rho\sigma}^{J}A_{\lambda}^{K} + \frac{3}{2}g
F_{\mu\nu}^{I}A_{\rho}^{J}(f_{LF}^{K}A_{\sigma}^{L}A_{\lambda}^{F})\right. 
\nonumber\\
&& +\left. 
\frac{3}{5}g^{2}(f_{GH}^{J}A_{\nu}^{G}A_{\rho}^{H})
(f_{LF}^{K}A_{\sigma}^{L}A_{\lambda}^{F})A_{\mu}^{I}\right\}-\frac14 a_{IJ}\cF^I_{\mu \nu}\cF^{J\mu \nu} \nonumber\\
&&+ g_{R} \left[ \frac{i}{2\sqrt{6}} \,
\bl^{i x} \l^{jy} \,P^{xy}_{ij} \, -\, \bl^{ix}
\g_{\mu} \psi^{j\mu}\, P^{x}_{ij} - \frac{i\sqrt{6}}{8} \,
\ol\psi^{i\mu} \g_{\mu \nu} \psi^{j\nu} \, P_{ij} \right] +
\nonumber\\
&&+ g  \left[2 \, \ol\psi^{i\mu} \g_{\mu} \z^A \, \cN_{iA} \, - 2 i \, \bz^A \l^{x i}\, \cM_{Aix}\,  +
 \bz^A \z^B \; \cM_{AB} \,  \right] \nonumber\\
&&- \frac12g_{\Sigma \Lambda} \cD_{\mu} \phi ^{\Sigma} \cD^{\mu} \phi ^{\Lambda}-V(\Phi)+ \nonumber\\
&&+ \cL_{Tensor}+\cL_{Pauli}+\cL_{4-fermions}+ \cL_{Others},
\end{eqnarray}
where the spin connection $\omega$ appears in the covariant derivative:
\beq
\nabla_{\mu}(\omega)=\left( \partial_{\mu} + \frac14 \omega_{\mu} {}^{ab}\gamma_{ab} \right) \nonumber
\eeq
and
\beq
\cD \phi^{x} = D \phi^{x} + g A^I K_I^{x} (\phi), \qquad \cD q^X = D q^X + g A^I K_I^X (q).\nonumber
\eeq
$K_I^{x}$ and $K_I^X$ are the Killing vectors for the gauging and 
\beq
L_I^{xy} \equiv \partial^{y} K_I^{x}, \qquad \omega_{I\,B}{}^A \equiv K_{IX;Y} f^{XA}_i f^{Yi}_B, \nonumber
\eeq
where $f_{iA}^X$ denote the vielbeins to pass the curved indexes $u$ to
the flat indexes $iA \in USp(2) \otimes USp(2n_H)$. ${\omega_{X\,i}}^j (q)$ and ${\Delta_{X\,A}}^B (q)$ are the
$USp(2)$ and $USp(2n_H)$ connections respectively for the quaternionic manifold $\cQ$. 
For what concerns the $\cS$ manifold, the completely symmetric constant tensor $C_{IJK}$, that appears in the
Chern-Simons couplings of the vector fields, determines also the geometry of the very special manifold. In fact,
$\cS$ can be described by an $(n_V +n_T)$--dimensional cubic hypersurface
\beq
C_{IJK} h^{I} h^{J} h^{K} = 1 \nonumber
\eeq
of an ambient space parametrized by $n_V + n_T +1$ coordinates $h^{I} = h^{I}(\phi^{x})$. 
$a_{IJ}$ are described in terms of $h^{I}$ and the metric $g_{xy}$:
\beq
a_{IJ} = h_{I} h_{J} + h^{x}_{I} h^{y}_{J}g_{xy}, \nonumber
\eeq
where 
\beq
\label{h}
h_{I,x} = \sqrt{\frac{2}{3}} h_{Ix}, \qquad h^{I}{}_{,x}=- \sqrt{\frac{2}{3}} h^{I}_{x}.
\eeq 
The Killing vector prepotential $P_{I\,i}{}^j$ depends only on $q^u$, while $P_{ij} \equiv h^I P_{Iij}$, $P^x_{ij} \equiv h^{xI}
P_{Iij}$ and $P^{xy}_{ij} \equiv \delta^{xy} P_{ij} + 4 T^{xyz}
P^{z}_{ij}$, where $T^{xyz}$ is a completely symmetric function of $\phi^x$, depend
on $q^u$ and $\phi^x$. From (\ref{h}), one obtains the following useful identity for us:
\beq
\label{relation}
P^{x}_{i j}= -\sqrt{\frac32} P^{,x}_{i j}
\eeq
The scalar potential $V(\Phi)$ is given by
\beq
\label{potential}
V = 2 g^2 W^{x} W^{x} - g^2_R \left[ 2
P_{ij} P^{ij} -P^{x}_{ij} P^{x\,ij}\right] + 2 g^2 \cN_{iA} \cN^{iA}.
\eeq 
The first contribution of (\ref{potential}) comes from tensor multiplets, while the rest part
comes from vector- and hyper-supermultiplets which under the condition (\ref{result}) gives (\ref{potstable}). 
Closure of the SUSY algebra leads: 
\beq
W^{x} = \frac{\sqrt{6}}{4} h^I K_I^{x},\qquad \cN_{iA} = \frac{\sqrt{6}}{4} f_{XAi} K^X_{I} h^{I}. \nonumber
\eeq
In the end, the mass matrices $\cM_{Aix}$, $\cM_{AB}$ are defined as
\beq
\cM_{Aix} \equiv f_{AiX} K^X_I h^{Ix}, \qquad \cM_{AB} \equiv \dfrac{i\sqrt{6}}{2} f_{AX}^i f_{BiY}
K^{[Y;X]}_I h^I. \nonumber
\eeq
More details on the notation and on the derivation of the full action and of the SUSY transformations can
be found in \cite{d5supergravity} and the references therein.

\end{document}